# On the history of photomultiplier tube invention

B.K.Lubsandorzhiev

*Institute for Nuclear Research of RAS*

Corresponding author:
Tel: +7(095)1353161; fax: +7(095)1352268;
e-mail: lubsand@pcbai10.inr.ruhep.ru
Address: 117312 Moscow Russia, pr-t 60-letiya Oktyabrya 7A
Institute for Nuclear Research of RAS

**Abstract**

In this very short note we review some historical aspects of photomultiplier tube invention. It is our tribute to the memory of great Soviet-Russian physicist and engineer *Leonid Aleksandrovitch Kubetsky* whose life and scientific achievements are described briefly. Particular efforts are made to shed light on a controversial issue of who invented the first photomultiplier tube. It is asserted that if to recognize L.A.Kubetsky's priority on the photomultiplier tube invention the last Beaune Conference would be held on the eve of the 75[th] Anniversary of that great event.

## I. Introduction

Photomultipliers (PMTs) are the most widespread vacuum electronic devices. Indeed, the PMTs are ubiquitous. The omnipresence of PMTs is striking. They are used practically in every kind of experimental studies including space research and archeology, medicine and geology, biology and art, astronomy and metallurgy, chemistry and agriculture, etc.

Physics experiments, particularly high energy physics and astroparticle physics experiments, are the most active users of PMTs. Moreover the most substantial achievements in PMT developments for the last 40 years have been made following requirements of physics experiments.

The question arises WHO INVENTED THE MARVELOUS DEVICE?

## II. The first PMT

On August 4 1930 Soviet-Russian physicist and engineer L.A.Kubetsky proposed new method and device to amplify weak photocurrents by dozens and hundreds thousands times without using traditional radiolamps [1]. In accordance with Kubetsky's proposal the device was furnished with a source of primary photoelectrons (photocathode) and consecutive secondary electron emitters (dynodes) with a certain coefficient $\sigma$ and each subsequent emitter was supplied with higher electrical potential in comparison with preceding emitter from high voltage power supply. Photoelectrons produced by the photocathode illumination were accelerated to the first emitter knocking out from it's surface secondary electrons. The latter are accelerated to the next emitter knocking out secondary electrons as well and so forth. The device can have arbitrary number of emitters. The electron flux from the last emitter are collected by collector or anode. The total gain of the device will amount to $\sigma^n$, where *n* is the number of emitters.

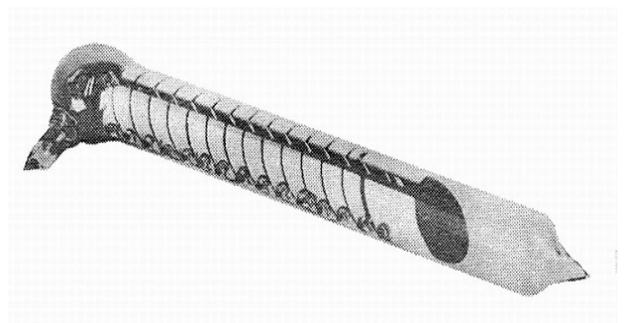

Fig.1. The first photomultiplier tube in the world: "Kubetsky's tube".

In 1933-34 L.A.Kubetsky has developed a number of photomultiplier tubes with Ag-O-Cs photocathodes and circular secondary electron emitters made also from Ag-O-Cs. The photomultiplier tubes consisted of photocathode and multistage electron multiplier system including constant magnets for electron focusing because electrostatic electron optics was not developed well at that time, Fig.1. The total gain of the tubes reached $10^3$-$10^4$ and more. Throughout 1930s the tubes have been renown in USSR as "Kubetsky's tubes" [2].

So the first photomultiplier tube was invented on August 4 1930 in Soviet Union by L.A.Kubetsky. It is "Kubetsky's tube". It is interesting to note that the last Beaune conference on New Developments in Photodetection has been held just one month before the 75$^{th}$ Anniversary of the PMT invention. It is a quite symbolic coincidence!

**IV. Controversy on the PMT invention authorship**

It is surprising that up to now the majority of physics community in the west has deepest conviction that the first PMT was developed by V.K.Zworykin et al. at RCA in 1936 [3]. There is practically no mentioning of L.A.Kubetsky's name in English scientific literature. As examples we make quotations from some articles.

*"It is difficult if not possible to, to find the origin of the idea of using cascaded stages of secondary emission to obtain electron current amplification. One very early name associated with this idea is that of Slepian of Westinghouse, who patented a multistage magnetic multiplier to serve as cathode in a cathode ray tube. Other names associated with this type of amplifier include P. T. Farnsworth, H. Iams and B.Salzbergs, G. Weiss etc. In 1934 a group working under V.K.Zvworykin at RCA undertook a serious study of this class of device, in particular because of its application to television. A paper was published in 1936 by Zvorykin, Morton and Malter discussing some of the early multipliers. It describes the first really successful magnetically focused multiplier …. "* [4].

*"… In 1936, Zworykin, Morton, and Malter, all of RCA reported on a multistage photomultiplier……."* [5]

*"… In the next year 1936, Zworykin et al. developed a photomultiplier tube having multiple dynode stages."* [6].

*"… the first photomultiplier tube was invented by the RCA laboratories in 1936 ……."* [7].

So far we encountered just with one unequivocal admission of Kubetsky's priority on the PMT invention in English scientific literature: *"…However up to 1930, when L.A.Kubetsky (1906-1959) basing on his author's certificate from 1930 made the first operating photoelectron multiplier, allowing to realize internal amplification of very weak electron fluxes by a factor of $10^3$-$10^6$, this idea has not been considered as neither realizable nor practically useful. Kubetsky's results have been picked up in USA (1934) and later in Germany and England"* [8] (It is translated by the author of the present paper back to English from the Russian edition of W.Summer's book [9]).

C.D'Ambrosio and H.Leutz in their beautiful review on Hybrid Photodiodes mentioned L.A.Kubetsky but not directly with regard to PMTs: *"… In 1936 and 1937 Zworykin et al. and Kubetsky achieved electron multiplication by secondary emissions from metal surfaces. "* [10].

So here we would like to highlight some historical facts, which are well known in Russia, e.g. see [11,12], to shed light on that controversy.

In September 1934 V.K.Zworykin was in USSR with business visit advertising RCA's products. He read lectures in Moscow and Leningrad. During that visit he attended L.A.Kubetsky's laboratory in Leningrad. L.A.Kubetsky demonstrated his tubes to the guest. The latter was very much impressed by the tubes and made certain of the tubes amplification of more than $10^3$ experimenting with the tubes himself [11-14]. There are a lot of evidences of Zworykin's visit to Kubetsky's laboratory and his experiments with Kubetsky's tube, e.g. see [11-14].

After that visit going back to USA V.K.Zworykin drew a sketch of photoelectron multiplier on a Berlin hotel paper. That sketch is dated by September 18 1934 and kept in David Sarnoff's archive [15]. As far as we know it is the first mentioning about photomultiplier tube in V.K.Zworykin's papers [12]. The first note concerning photomultiplier tube in V.K.Zworykin's laboratory journal is dated by November 22 1934 [15]. At last, in 1936 V.K.Zworykin and his colleagues published their famous paper [3] which is considered in the west as a starting point of the PMT history. Even a fleeting glance at the Zworykin's and Kubetsky's tubes reveals their great similarities: the same magnetic focusing etc.

**III Leonid Aleksandrovitch Kubetsky**

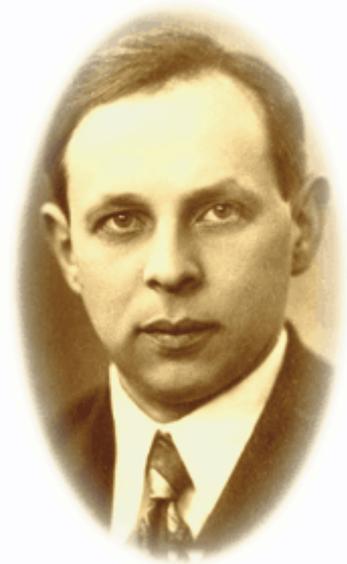

Fig.2. Leonid Aleksandrovitch Kubetsky (1906-1959).

Leonid Aleksandrovitch Kubetsky was born on July 12 1906 in Tsarskoye Selo (now Pushkin Leningrad region) in the poor family of clergyman. His childhood coincided with very hard times in Russian history and was full of hardships and deprivations. Nevertheless he finished high school brilliantly with recommendations to be enlisted to any university without competitive examinations for "outstanding achievements in studies".

In 1923 he started to study physics at the faculty of physics and mathematics of Petrograd University (now S-Peterburg State University). At that time he was forced to combine his studies at University with working as an electrician to earn his living. He was not fully satisfied with a formal style of teaching at the University and in 1925 he moved to the electromechanical faculty of famous Leningrad Politechnical Institute continuing to work as an electrician.

As a student he started to work in the technical department of Physico-Technical Rentgen Institute in the field of a low current electrical engineering. In 1929 he developed a gas discharge device with heated cathode and controlled electrode [16]. In the next 1930 being a student he proposed a photoelectron multiplier. As practically every genius stroke it was put forward by L.A.Kubetsky when he was just 24 years old and a student still! Almost at the same time he developed "Cathode transmitter for television" [17] and "Optical microphone" [18]. The former was a prototype of transmitting TV tube proposed a bit later by P.T.Farnsworth.

In 1931 L.A.Kubetsky graduated from Institute and started to work at Leningrad Electro-Physical Institute. In the following years L.A.Kubetsky has worked at a number of physics institutes in Leningrad and Moscow refining his photomulptiplier tube, developing electron multiplier systems, new vacuum devices etc, for more information see [11].

Particular emphasis he has made on photomultiplier tubes application. In 1939 he discovered infrared emission of the night sky [19]. He made substantial contribution to the development of dissectors, iconoscopes and other kinds of vacuum tubes. He proposed new methods of ultracontrast conversion of spectra and images, ultracontrast fine structure analysis etc.

Unfortunately the last years of his life have been darkened by problems with his health in particular after a serious surgical operation carried out in 1948. L.A.Kubetsky died on September 22 1959. He was just 53. He was buried in Vostryakovskoye cemetery in Mocsow.

## IV. Conclusion

Least of all we would like to shake authority of V.K.Zworykin (by the way he was Russian by origin too) or to offend his memory. We have no right to judge and we can not do it. We will never know why L.A.Kubetsky's name was buried in oblivion in the west. But after all the world and time changed. The world is not divided by iron curtain anymore. Now it is time to restore historical justice and to pay tribute to a great physicist and engineer Leonid Aleksandrovitch Kubetsky. It is time now to admit his priority on the PMT invention at last.

Anyway his life was rewarded by flourishing of new fields of scientific studies and experimental techniques which became possible due to L.A.Kubetsky's brainchild: wonderful magic vacuum electronic device – PHOTOMULTIPLIER TUBE!

**Acknowledgements**

The author would like to devote this paper to the memory of his teacher B.M.Glukhovskoy. Almost twenty years ago he was going to write a paper on the PMT history and publish it in an international journal. He didn't manage to do it because of untimely decease. It is my humble attempt to accomplish my teacher's desire. I would like to thank my colleagues from MELZ Company in Moscow for many invaluable information on the PMT history and Dr.V.Ch.Lubsandorzhieva for careful reading the paper and many valuable remarks.

**Figures:**

Fig.1. The first photomultiplier in the world: "Kubetsky's tube".

Fig.2. Leonid Aleksandrovitch Kubetsky (1906-1959).